\documentclass[12pt]{iopart}
\usepackage{graphicx}
\usepackage{color}
\usepackage{umlaut}
\usepackage[english]{babel}

\begin{document}
\title{Temperature dependence of the energy dissipation in dynamic force microscopy}

\author{Tino Roll, Tobias Kunstmann, Markus Fendrich, Rolf Möller, Marika Schleberger \footnote{electronic address: marika.schleberger@uni-due.de}} 
\address{Fachbereich Physik, Universität Duisburg-Essen, D-47048 Duisburg, Germany}

\begin{abstract}
The dissipation of energy in dynamic force microscopy is usually described in terms of an adhesion hysteresis mechanism. This mechanism should become less efficient with increasing temperature. To verify this prediction we have measured topography and dissipation data with dynamic force microscopy in the temperature range from 100 K up to 300 K. We used 3,4,9,10-perylenetetracarboxylic-dianhydride (PTCDA) grown on KBr(001), both materials exhibiting a strong dissipation signal at large frequency shifts. At room temperature, the energy dissipated into the sample (or tip) is 1.9~eV/cycle for PTCDA and \mbox{2.7~eV/cycle} for KBr, respectively, and is in good agreement with an adhesion hysteresis mechanism. The energy dissipation over the PTCDA surface decreases with increasing temperature yielding a negative temperature coefficient. For the KBr substrate, we find the opposite behaviour: an increase of dissipated energy with increasing temperature. While the negative temperature coefficient in case of PTCDA agrees rather well with the adhesion hysteresis model, the positive slope found for KBr points to a hitherto unknown dissipation mechanism.
\end{abstract}

\maketitle

\section{Introduction}
\label{sec:Introduction}
Non-contact scanning force microscopy presents a unique tool to study surfaces even on the atomic scale \cite{Giessibl1995}. In contrast to scanning tunneling microscopy, imaging is not restricted to conducting samples. In addition to the well known capability to produce atomically resolved topgraphical images, the method can be used to study energy loss processes as well. This can be achieved by measuring the energy necessary to maintain the amplitude of the cantilever (damping) at a given frequency shift $\Delta f$. This {\it dissipation} data does not necessarily correspond directly to the topography and in many cases atomic-scale contrast can be obtained, often revealing additional information. However, a thorough understanding of the underlying processes has not yet been achieved. To explain the experimentally observed dissipation images, several mechanisms have been suggested (see e.g. \cite{moritabchap19-20}), two of them being considered to be the most likely ones. 

One is based on an adhesion hysteresis effect \cite{Sasaki,Kantorovich04PRL93_236102}, the other is connected to the stochastic friction the tip experiences in the vicinity of the surface due to the vibrations of the atoms in the tip-surface junction \cite{Gauthier1999}. While the latter predicts dissipated energies which are too small by orders of magnitude, the first model predicts dissipated energies of 0.01 eV - 2~eV which are of the order of characteristic binding energies and agree well with the experimental data. In this model, the cause for the significant dissipation is a double-welled potential energy surface. When the tip approaches the surfaces very closely a reversible structural change in the tip surface-junction like a jumping surface atom or a flipping surface fragment may occur. Upon retraction, the surface atom (or fragment) returns to its original position. Thus, over every oscillation cycle the tip experiences a hysteresis in the acting forces resulting in a detectable energy loss. Recently, good agreement between experimental data and theoretical models could be shown for Ge(111)C(2x8)\cite{Oyabu2006}.

It is obvious that such a mechanism should depend strongly on the temperature. The surface atom may overcome the potential barrier of the double well potential more easily if the temperature is raised and thus its thermal energy increases. Therefore, with increasing temperature the hysteresis becomes smaller and the energy dissipation should decrease. In contrast, the stochastic friction model predicts just the opposite: The friction caused by the surface atoms basically depends on their velocity and should increase with increasing temperature. To our knowledge, so far no experimental data exists on this topic. 

In this paper, we study for the first time the dissipated energy as a function of temperature. As samples we used crystallites of the organic molecule PTCDA grown on the insulator KBr(001). This heterosystem is well studied \cite{Nony2004,Nony2004NanoLett,Mativetsky,Kunstmann2005} and allows the direct comparison of the dissipation signal from an organic molecule with the signal obtained from an ionic surface, as the organic molecules show island growth and do not form a wetting layer on KBr(001). 

\section{Experimental}
\label{sec:Experimental}
All experiments have been performed in an ultra high vacuum (UHV) system at variable temperatures. To cool down the sample we have used a liquid helium flow cryostat. To avoid transmissions of acoustic noise and to dampen the cryostat vibrations, the sample stage is coupled to the cryostat via soft copper foil stripes. In addition, the sample stage is suspended with springs and damped by an eddy current setup. The measurements were performed at 300~K, 200~K, 180~K and 100~K. To obtain a stable temperature, the cooling was counterbalanced by a radiative heater. The temperature can be measured with a built-in silicon diode and a thermocouple which is mounted directly to the sample. The absolute accuracy of the temperature is thus better than 1~K. The cantilever and the sample reached an equilibrium temperature after approximately two hours. 

The scanning probe microscope (RHK AFM/STM UHV 7500) was operated in the frequency modulation detection mode \cite{Albrecht91JAP69_668}. One feedback loop controls the separation between the tip and the sample by keeping the frequency shift $\Delta f$ at a certain value. This signal contains the topographical information. The second feedback controller keeps the oscillating amplitude constant at about $A=34$~nm by adjusting the drive amplitude. By measuring this quantity we can determine the average energy dissipation related to the tip-sample interaction. Prior to the experiments, a bias voltage of typically 0.2~V has been applied to compensate for any long ranged electrostatic interactions between tip and sample. All measurements have been performed with the same cantilever (QNCHR by Nanosensors, specifications: $f_0\approx$~310~kHz, $Q\approx$~18000, $k\approx$~42~N/m). 

The KBr(001) surface was prepared by {\it ex situ} cleaving followed by immediate introduction into the load lock of the vacuum chamber. In order to remove contaminations and trapped charges, the load-lock chamber was heated at 400 K for 3 hours. During heating, the base pressure in the load lock was $1.1\times10^{-7}$mbar. Subsequently the substrate was transferred to the preparation chamber where the PTCDA was thermally evaporated from a home built crucible at a base pressure of $\leq1\times10^{-8}$mbar. The sample was kept at room temperature during the evaporation. The base pressure during AFM measurements was $\leq3\times10^{-10}$mbar. XPMPro software \cite{xpmpro} was used for data acquisition as well as for image processing. Except for contrast enhancement in the case of fig.~\ref{figure3} and for plane subtraction, all images shown are raw data.

\section{Results}
\label{sec:Results}

\begin{figure}[htb]
\includegraphics[width=15.0cm]{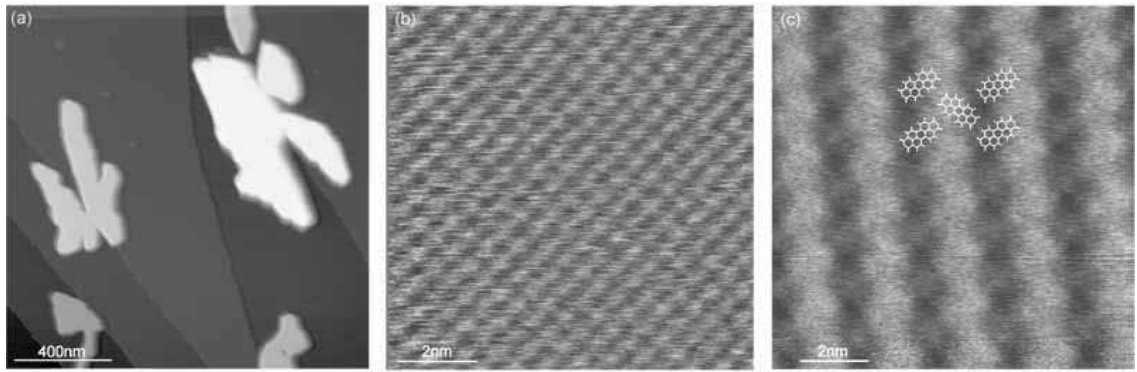}
\caption{(a) Topography of PTCDA crystallites grown on KBr(001). Image has been acquired using $\gamma = -14$~fNm$^{1/2}$. Height range is about 6~nm. (b) High resolution image of the uncovered KBr(001) surface. Normalized frequency shift is $\gamma = -16$~fNm$^{1/2}$. (c) Molecular resolution of PTCDA obtained on the top of a crystallite. Here, the normalized frequency shift is $\gamma = -20$~fNm$^{1/2}$. Individual PTCDA molecules in the typical herringbone structure can be resolved in the topography signal. All images have been acquired at 300~K.}
\label{figure1}
\end{figure}

Fig.~\ref{figure1}(a) gives an overview of a large-area scan at room temperature of the flat KBr(001) surface covered with PTCDA crystallites. The (median) height of these islands is 2-3 nm. Zooming in, atomic resolution on the substrate (fig.~\ref{figure1}(b)), as well as molecular resolution on the topmost molecular layer (fig.~\ref{figure1}(c)) could be achieved. The normalized frequency shift \cite{Giessibl1995} is $\gamma = \frac{\Delta f}{f_0} k A^{3/2} =-14$ fNm$^{1/2}$ for the large area scan, $\gamma = -16$ fNm$^{1/2}$ for the high resolution image of KBr(001) and $\gamma = -20$ fNm$^{1/2}$ for the molecular resolution image. The crystallite growth seems to nucleate from the step edges of the KBr substrate and the growth mode is clearly of the Volmer-Weber type, resulting in PTCDA crystallites on top of a clean KBr substrate without any wetting layer. These results are in good agreement with former experiments \cite{Kunstmann2005,Fendrich2007} and show that the sample is well prepared.

\begin{figure}[htb]
\includegraphics[width=16.5cm]{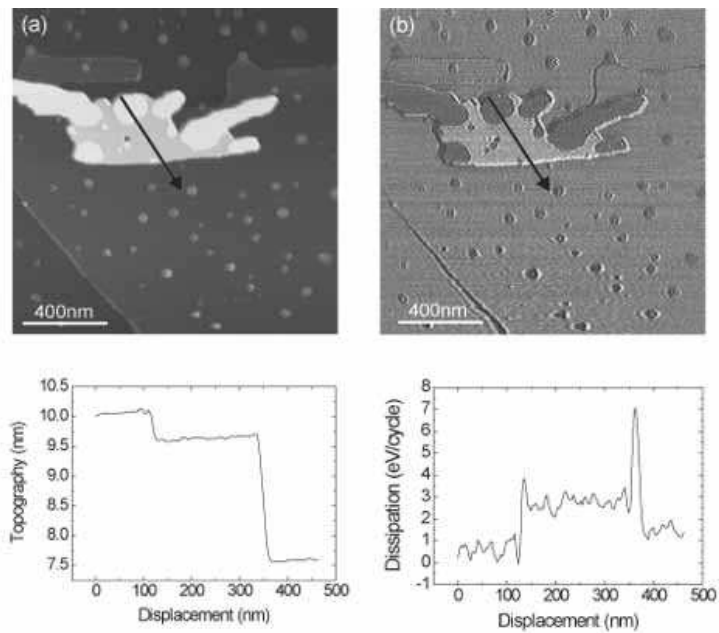}
\caption{(a) Topography of PTCDA and KBr(001) taken at 180~K. Image has been acquired using $\gamma = -14$~fNm$^{1/2}$. Height of the PTCDA crystallite is about 2~nm. (b) The corresponding dissipation image exhibits a clearly enhanced energy loss of the cantilever which oscillates above the PTCDA crystallite. Dissipation due to tip-sample interaction is about 0.5~eV/cycle when scanning over areas covered with residual adsorbates. Oscillating above the PTCDA and KBr(001) regions lead to a dissipation of 2.5~eV/cycle and 1.5~eV/cycle, respectively.}
\label{figure2}
\end{figure}

For the investigation of the dissipation behaviour of KBr and PTCDA as a function of temperature, we have performed large area scans with a typical frame size of 1.5 $\times$ 1.5 $\mu$m$^2$. To assure the same experimental conditions, the dissipation at a given temperature was measured for PTCDA and KBr in the same run. Because pronounced topographic features such as step edges are known to give rise to artefacts in dissipation images, both scanning directions and the error signals of the two control loops were checked carefully. A careful examination of the topography images yielded no conspicious events at the step edges thus giving no indication for possible tip changes. This can e.g. be seen from the line scan shown in fig. \ref{figure2}. Therefore, we believe the tip to be stable within the same measurement. The $df$ signal and the oscillation amplitude were recorded simultanuously. The latter was checked to be constant. The typical scanning velocity was about 750~nm/s. 

In order to determine the dissipated energy due to tip-sample interactions correctly one has to take the intrinsic dissipation of the lever due to internal friction into account. Therefore, we calculated the intrinsic dissipated energy of the freely oscillating cantilver according to 
\cite{Anczykowski1999} 
\begin{equation}
	E_{0}= \frac{\pi kA^{2}}{Q}.
\end{equation}
For the cantilever oscillating at its free eigenfrequency we find a energy dissipation of about about~31~eV/cycle at all temperatures with the exception of the measurement at 200~K. Here, the internal dissipation was marginally lower. 
Due to noise during the measurements and the accuracy of measuring the quality factor $Q$, all values of dissipation given in this paper have an absolute  error of about 25\%. At any given temperature, the value of the dissipated energy is constant within 2\%. Hence, we take this as an indication that the dissipation signal is not masked by any instrumental artifacts.

Figure \ref{figure2} shows typical topography (a) and dissipation images (b) taken at 180~K. The corresponding linescans are also given. The direction of the linescans is indicated by arrows. The apparent height of the PTCDA crystallite is about 2~nm. Because of the cryogenic effect of the cooled sample, residual adsorbates are found close to the edges of the PTCDA crystallite and on the KBr surface. 
In the dissipation image we are able to identify three different regions. As one can see from the linescan in \ref{figure2}(b), in areas of the adsorbate layer the dissipated energy is lower than in the KBr and PTCDA regions, respectively. In comparison, the dissipation signal is clearly largest while scanning above the PTCDA crystallite.

By analyzing the line scans one can determine the dissipated energy \cite{Anczykowski1999}, where $V_{exc}$ and $V_{exc,0}$ are the voltages needed to maintain the amplitude close and far away from the surface, respectively: 
\begin{equation}
	E_{ts}=E_{0}\left(\frac{V_{exc}}{V_{exc,0}}-\frac{f}{f_0}\right)
\end{equation}
As one can see from the linescan in fig.~\ref{figure2}(b), the totally dissipated energy during one oscillation cycle was 1.5~eV/cycle at the substrate areas and  2.5~eV/cycle at the PTCDA crystallite.

\begin{figure}[htb]
\includegraphics[width=11.5cm]{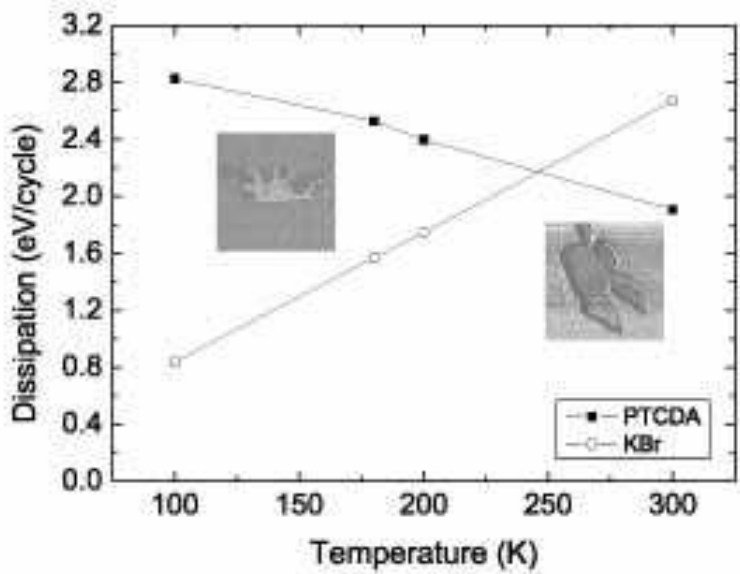}
\caption{Dissipated energy per oscillation cycle as a function of the sample temperature. Squares refer to the dissipation signal of the PTCDA, open circles represent the signal from the KBr(001) substrate. Lines are drawn to guide the eye. Data points at a given temperature are taken during the same measurement session. Insets show contrast enhanced dissipation data at 180 and 300~K, respectively. Normalized frequency shift is $\gamma = -14$~fNm$^{1/2}$.}
\label{figure3}
\end{figure}

To determine the mean dissipated energy in a more accurate way, we selected areas which correspond to the KBr substrate and the PTCDA, respectively. Within these areas we determined the mean dissipated energy per oscillation cycle by averaging over the respective areas in the dissipation image. Fig.~\ref{figure3} shows the development of the energy dissipation for KBr and PTCDA as a function of the sample temperature. Two respective data points taken at a given temperature are acquired during the same measurement. Therefore, we can exclude any changes in terms of the experimental setup such as e.g. base pressure, sample contamination or tip changes. In order to verify the control parameters of the self oscillation mode, the oscillation amplitude has been recorded for each scan. By keeping the amplitude constant we assured that the feedback loops were adjusted correctly and any contrast in the dissipation channel therefore corresponds to non-conservative damping mechanisms only. 

The data in fig.~\ref{figure3} exhibits a striking difference in the slope. Let us consider the characteristics of the PTCDA dissipation signal first. At room temperature, we find that the average dissipated energy amounts to 1.9~eV/cycle. In a recent study of PTCDA grown on KBr \cite{Fendrich2007} an energy dissipation of 0.9~eV/cycle was found for a normalized frequeny shift of $\gamma=-8$ fNm$^{1/2}$ and 0.6~eV/cycle for $\gamma=-4$ fNm$^{1/2}$, respectively. That is, considering that our data was taken at $\gamma=-14$ fNm$^{1/2}$, the value we find at room temperature is in good agreement with the moleculary resolved data from ref. \cite{Fendrich2007}.  From the data shown in fig.~\ref{figure3}, a nearly linear temperature dependence becomes apparent. The signal drops from 2.8~eV/cycle at 100~K to about 1.9~eV/cycle at 300~K. This means that by increasing the  sample temperature by 100~K the dissipated energy per oscillation cycle decreases by an amount of 0.45~eV/cycle. 

Taking into account the amount of energy dissipation and the temperature dependence found, there is only one mechanism capable of explaining our data. The adhesion hysteresis model postulates that a sudden change in the atomic configuration in the tip-surface junction leading to a double-welled potential. This again gives rise to a hysteresis in the tip-surface force and thus to energy dissipation \cite{Kantorovich04PRL93_236102,Kantorovich2005}. This mechanism has just recently been clearly identified in the case of a semiconducting surface \cite{Oyabu2006} and other experimental results yield the predicted amount of energy dissipation \cite{Schirmeisen05}. The negative coefficient, i.e. as the temperature is increased, the energy disspation decreases, is due to the fact that the surface (or tip) atom is able to cross the potential energy barrier earlier on approach and retraction. Simulations \cite{Kantorovich04PRL93_236102} show in principle the same quantitative behaviour as determined in our experiments. 

Obviously, the slope of the dissipation measured on the KBr region shown in fig.~\ref{figure3} cannot be explained by the mechanism described above. With decreasing temperature it should become more and more difficult for the surface (or tip) atom to overcome the potential barrier resulting in a larger hysteresis effect and an increasing dissipation. Instead, we find that the dissipated energy decreases nearly linearly from 2.7~eV/cycle at 300~K to about 0.8~eV/cycle at 100~K. If the sample temperature is raised by 100~K, the dissipated energy decreases by an amount of 0.95~eV/cycle. That is, the temperature coefficient for the KBr substrate is not only negative but also twice as large as the one found for the PTCDA crystallite. 

Dissipation values measured at RT at KBr and NaCl surfaces, respectively, are
lower than the value measured here by an order of magnitude \cite{Schirmeisen2006,Hoffmann2007}. A direct comparison is, however, difficult because the absolute value depends strongly on parameters such as $df$ or the amplitude. In addition, our experiment was performed with a heterosystem, so the tip might have picked up a PTCDA molecule giving rise to a rather large dissipation signal. Since we are confident that the tip is stable during each measurement this would however not change the observed trends. At 8~K Hoffman et al.~found a value of only 0.01~meV/cycle (measured at a comparable normalized frequency shift of $\gamma=-2~$fNm$^{1/2}$ \cite{Hoffmann2007}), which is not in contradiction to our data if the observed temperature trend is extrapolated linearly to extremely low temperatures.

If we assume that a stochastic friction force mechanism as first suggested in \cite{Gauthier1999} is the underlying process for the energy dissipation in the KBr region, the positive temperature coefficient, i.e. the increase of dissipation with increasing temperature, could be accounted for. In this model the tip experiences a frictional force due to the random vibrations of atoms in the surface and tip - just like a brownian particle immersed in a fluid of light particles. This force is proportional to the thermal motion of the surface atoms. Increasing the temperature, the thermal motion is enhanced and the friction should increase as well. However, the calculated magnitude of energy dissipation due to the stochastic friction force mechanism is at least 10$^{5}$ times smaller \cite{Gauthier1999,Trevethan2004,KantorovichPhys2004} than observed in our experiments. Thus, despite giving the right trend for the observed temperature coefficient, the friction force mechanism in its current form cannot be responsible for the dissipation measured here. 

Note that both models aim to explain the atomic scale contrast found in dissipation images. Nevertheless, by scanning a larger frame, the tip samples over many atomic sites and the corresponding average dissipation should show at least the same trend as predicted for the atomic scale.

\section{Conclusion}
\label{sec:Conclusion}
We have presented first experimental data on the temperature dependence of energy dissipation in dynamic force microscopy. For crystallites of the organic molecule PTCDA, we find the theoretical predictions given by the adhesion hysteresis model confirmed. The amount of dissipated energy as well as the temperature coefficient are in good agreement with the theory. For the ionic crystal KBr, the magnitude of energy dissipation is also well within the predicted range, however, the temperature dependence is in clear contradiction to the expectations from the model. Since the only model predicting the right temperature behaviour cannot account for the amount of dissipated energy, we conclude that there must exist another dissipation mechanism not yet described by theory. Note, that in the current experiment the tip is at room temperature while the sample is at low temperature which could open a dissipation channel not yet considered. Thus, further development of theoretical models for the energy dissipation in dynamic force microscopy is needed.

\section{Acknowledgement}
\label{sec:Acknowledgement}

The authors like to thank M. Klocke and D. Wolf for fruitful discussions.
Financial support from the Deutsche Forschungsgemeinschaft through SFB 616 "Energy dissipation at surfaces" is gratefully acknowledged.

\section*{References}

\end{document}